\documentclass[12pt]{article}
\begin{document}
\title{A Note on New Fine Structure Energy Levels}
\author{B.G. Sidharth\\
G.P. Birla Observatory \& Astronomical Research Centre\\
B.M. Birla Science Centre, Adarsh Nagar, Hyderabad - 500 063
(India)}
\date{}
\maketitle

\begin{abstract}
We consider a slight modification of the infinite range Coulomb potential to take account of the finite size of the universe. Surprisingly, the energy levels are almost doubled.
\end{abstract}
\begin{flushleft}
\end{flushleft}
We start by factoring in the finite size of the universe in the Coulomb potential which has infinite range. This can be done if we introduce the Yukawa potential,
\begin{equation}
V(r) = \alpha e^{-\mu r}/r\label{4ebb1}
\end{equation}
in place of the Coulomb potential. This has the range $r \sim \frac{1}{\mu}$ which can be taken to be the radius of the universe,
into the Dirac equation instead of the usual Coulomb potential
\cite{greinermuller}.  We introduce (\ref{4ebb1}) into the stationary Dirac
equation to get
\begin{equation}
[c \vec{\alpha} \cdot \vec{p} + \beta m_0c^2 - (E - V(r))] \psi (r)
= 0\label{4ebb2}
\end{equation}
From (\ref{4ebb2}) and (\ref{4ebb1}), we can immediately see that
roughly the effect is to shift the energy levels by a miniscule amount.\\
We further introduce the notation
\begin{equation}
Q = 2 \lambda r, \quad \mbox{where} \, \lambda = \frac{\sqrt{m^2_0
c^4 - E^2}}{hc},\label{4ebb3}
\end{equation}
After the standard substitutions (Cf.ref.\cite{greinermuller}) we
finally obtain
$$\frac{d\Phi_1}{dQ} = \left(1 - \frac{\alpha E}{hc\lambda Q}\right) \Phi_1 - \left(\frac{\chi}{Q} + \frac{\alpha m_0c^2}{hc\lambda Q}\right) \Phi_2,$$
\begin{equation}
\frac{d\Phi_2}{dQ} = \left(- \frac{\chi}{Q} + \frac{\alpha
m_0c^2}{hc\lambda Q}\right) \Phi_1 + \left(\frac{\alpha E}{hc\lambda
Q}\right) \Phi_2\label{4ebb4}
\end{equation}
The substitutions
$$\Phi_1 (Q) = Q^\gamma \sum^{\infty}_{m=0} \alpha_m Q^m,$$
\begin{equation}
\Phi_2 (Q) = Q^\gamma \sum^{\infty}_{m=0} \beta_m Q^m.\label{4ebb5}
\end{equation}
in (\ref{4ebb4}) leads to
$$\alpha_m (m+\gamma) = \alpha_{m-1} - \left(\frac{\alpha E}{hc\lambda}\right) \alpha_m -
\left(\chi + \frac{\alpha m_0c^2}{hc\lambda}\right)\beta_m,$$
\begin{equation}
\beta_m (m+\gamma) = \left(-\chi + \frac{\alpha
m_0c^2}{hc\lambda}\right) \alpha_m + \left(\frac{\alpha
E}{hc\lambda}\right)\beta_m.\label{4ebb6}
\end{equation}
As is well known $\gamma$ in (\ref{4ebb5}) is given by
\begin{equation}
\gamma = \pm \sqrt{\chi^2 - \alpha^2},\label{4ebb7}
\end{equation}
where $\lambda$ is given by (\ref{4ebb3}).\\
At this stage we remark that the usual method adopted for the
Coulomb potential is no longer valid - mathematically, Sommerfeld's
polynomial method becomes very complicated and even does not work
for a general potential: We have to depart from the usual procedure
for the Coulomb potential in view of the Yukawa potential
(\ref{4ebb1}). Nevertheless, it is possible to get an idea of the
solution by a slight modification. This time we have from
(\ref{4ebb6}), instead the equations
$$\alpha_m (m+\gamma) = \alpha_{m-1} - \frac{\alpha E}{\hbar c\lambda} \alpha_m +
\frac{\alpha E \mu}{\hbar c\lambda} \alpha_{m-1}$$
\begin{equation}
(-\chi \beta_m) - \frac{\alpha m_oc^2}{\hbar c\lambda} \beta_m +
\frac{\mu \alpha m_0c^2}{\hbar c\lambda} \beta_{m-1}\label{4ebb8}
\end{equation}
and
$$(m + \gamma) \beta_m = - \left(\chi + \frac{\alpha m_0c^2}{\hbar c \lambda}\right)
\alpha_m - \frac{\mu \alpha m_0c^2}{\hbar c\lambda} \alpha_{m-1}$$
\begin{equation}
+ \frac{\alpha E}{\hbar c \lambda} \beta_m - \frac{\mu \alpha
E}{\hbar c \lambda}  \beta_{m-1}\label{4ebb9}
\end{equation}
After some algebra on (\ref{4ebb8}) and (\ref{4ebb9}) we obtain
\begin{equation}
P \alpha_m + Q \beta_m = R \alpha_{m-1}\label{4ebb10}
\end{equation}
\begin{equation}
S \alpha_m + T \beta_m = U \beta_{m-1}\label{4ebb11}
\end{equation}
where $P,Q,S,T$ can be easily characterized, in the derivation of
which we will neglect
$\mu^2$ and higher orders.\\
If
$$\alpha_m / \beta_m \equiv p_m$$
then we have from (\ref{4ebb10}) and (\ref{4ebb11}),
$$S p_m + T = \frac{U (QS-PT)}{(RSp_{m-1} - PU)}$$
We note that the asymptotic form of the series in (\ref{4ebb5}) will
not differ much from the Coulomb case and so we need to truncate
these series. For the truncation of the series we require
$$QS = PT$$
This gives
$$\left\{1 + \frac{\chi \hbar c \lambda}{\alpha m_0c^2}\right\} \left[E\left(\chi +
\frac{\alpha m_0c^2}{\hbar c \lambda}\right) + m_0c^2 \left(m+\gamma
- \frac{\alpha E} {\hbar c \lambda}\right)\right]$$
$$ = \left(m+\gamma - \frac{\alpha E}{\hbar c \lambda}\right) \frac{\hbar c \lambda}
{\alpha m_0c^2} \left[E\left(m+\gamma + \frac{\alpha E}{\hbar c
\lambda}\right) + m_0c^2 \left(\chi + \frac{\alpha m_0c^2}{\hbar
c\lambda}\right)\right]$$
$(m = 1,2,\cdots)$\\
Further simplification yields
$$(m+\gamma)^2 + \left\{\frac{\chi \hbar c \lambda + \alpha m_0c^2}{\hbar c \lambda}
\right\}^2 + \frac{\alpha^2 E^2}{\hbar^2c^2\lambda^2}$$ where
$\gamma$ is given by (\ref{4ebb7}). Finally we get, in this
approximation,
\begin{equation}
E^2 = m_0^2 c^4 \left[1 - \frac{2\alpha^2}{\alpha^2 + (m+\gamma)^2 -
\chi^2}\right] + A O(\frac{1}{m^2})\label{4ebb12}
\end{equation}
In (\ref{4ebb12}) $A$ is a small quantity
$$A \sim m^2_0 c^4 - E^2$$
The second term in (\ref{4ebb12}) is a small shift from the usual
Coulombic energy levels. In (\ref{4ebb12}) $m$ is a positive integer
and this immediately provides a comparison with known fine structure
energy levels. To see this further let us consider large values of
$m$. (\ref{4ebb12}) then becomes
\begin{equation}
E = m_0c^2 \left[1 - \frac{\alpha^2}{m^2_1}\right]\label{4ebb13}
\end{equation}
while the usual levels are given by
\begin{equation}
E = m_0c^2 \left[ 1 - \frac{\alpha^2}{2m^2_2}\right]\label{4ebb14}
\end{equation}
$m_1$ and $m_2$ position integers.\\
We can see from (\ref{4ebb13}) and (\ref{4ebb14}) that the Yukawa potential
reproduces all the energy levels of the Coulomb potential but
interestingly (\ref{4ebb13}) shows that there are new energy levels
because $m_1^2$ in the new formula can be odd or even but $2m^2_2$
in the old formula is even. However all the old energy levels are
reproduced whenever $m^2_1 = 2m^2_2$ that is, even if $\mu = 0$, then as can be
seen from (\ref{4ebb2}), we get back the Coulomb problem. In any
case, the above calculation indicates how the problem changes.\\
We must remember that these fine structure energy levels like the Lamb shift which is of the order of $30 cm$ wavelength are in the Radio range. So the effect of these new levels would be minute signals in the radio spectrum beyond the well known $21 cm$ region of Hydrogen.\\
As is well known, these radio waves are reflected off the earth's ionosphere and would be difficult to detect on the earth. But recently NASA's balloon borne ARCADE experiments, operating from some 40 kms above the earth's surface, have tuned in to precisely such a radio sea \cite{kogut,bgsijmpe}.\\
Finally, it may be pointed out that all this could be interpreted alternatively as the photon having a miniscule mass $\sim 10^{-64} gm$ \cite{tduniv}. In this case also the above analysis would apply \cite{tduniv,fpl,ijmpe}

\end{document}